# Torsional instability in the single-chain limit of a transition metal trichalcogenide

Thang Pham[1,2,3,4], Sehoon Oh[1,2], Patrick Stetz[1,2,4], Seita Onishi[1,2,4], Christian Kisielowski[5], Marvin L. Cohen[1,2], and Alex Zettl[1,2,4]*

[1]Department of Physics, University of California at Berkeley, Berkeley, CA 94720, USA
[2]Materials Sciences Division, Lawrence Berkeley National Laboratory, Berkeley, CA 94720, USA
[3]Department of Materials Science and Engineering, University of California at Berkeley, Berkeley, CA 94720, USA
[4]Kavli Energy NanoSciences Institute at the University of California at Berkeley, Berkeley, CA 94720, USA
[5]The Molecular Foundry and Joint Center for Artificial Photosynthesis, One Cyclotron Road, Berkeley California 94720 USA

*Corresponding author: azettl@berkeley.edu

**One sentence summary:** Experimental and theoretical investigation demonstrates that, in the single- to few-chain limit, the van der Waals bonded linear chain conductor $NbSe_3$ is unstable towards a new form of dynamic charge-induced torsional wave (CTW).

**Abstract**:

We report the synthesis of the quasi-one-dimensional transition metal trichalcogenide $NbSe_3$ in the few chain limit, including the realization of isolated single chains. The chains are encapsulated in protective boron nitride or carbon nanotube sheaths to prevent oxidation and to facilitate characterization. Transmission electron microscopy reveals static and dynamic structural torsional waves not found in bulk $NbSe_3$ crystals. Electronic structure calculations indicate that charge transfer drives the torsional wave instability. Very little covalent **bonding is found between the chains and the nanotube sheath, leading to relatively unhindered longitudinal and torsional dynamics for the encapsulated chains.**

The successful isolation of monolayers of van der Waals bonded quasi-two-dimensional solids such as graphite (*1*) and the transition metal dichalcogenides (TMDs) (*2*) has spurred intense renewed experimental and theoretical interest in these low dimensional materials. Monolayer or few-layer sheets of graphene or TMDs often display dramatically different electronic, optical, and structural properties from those of the bulk materials, with profound underlying physics and far-ranging applications potential.(*3*) Transition metal trichalcogenides (TMTs) such as $NbSe_3$ and $TaS_3$ are closely related van der Waals bonded quasi-*one*-dimensional linear chain compounds which have been previously extensively studied in bulk form. These materials can support unusual ground states and collective-mode electronic transport.(*4*) Although some attempts have been made to study thinned TMTs (e.g. $NbSe_3$ samples have been cleaved down to ~200 chains in width (*5*)), no experimental or theoretical study has examined TMTs in the single- or few-chain limit. It is far more difficult to isolate and manipulate atomic chains than atomic sheets, and atomically thin samples can be highly air sensitive.(*6*)

Here we present a facile and effective method to prepare low-number chains of $NbSe_3$ within carbon and boron nitride nanotubes (CNTs and BNNTs). The spatial confinement promotes and stabilizes the growth of sub-unit cell $NbSe_3$ down to triple, double and even single atomic chains. Encapsulation additionally protects the chains from environmental oxidation and facilitates easy handling and characterization. The chains are mobile within the tubes. Unusual helical torsional waves with regular periodicity are observed, even in the single chain limit. Complementary theoretical calculations show that the electronic band structure of NbSe$_3$ is highly dependent on chain number and orientation, and that the torsional wave instability is driven by charging of the chains. We term this new phenomenon a charge-induced torsional wave (CTW).

NbSe$_3$ chains are directly grown via vapor transport inside the hollow cavity of pre-formed and open-ended multiwall CNTs and BNNTs (see Materials and Methods (*7*)). Related techniques have been previously used to encapsulate foreign species within nanotubes.(*8−15*) Once the synthesis is complete, the samples can be exposed to air and liquids with no apparent degradation of the encapsulated chains.

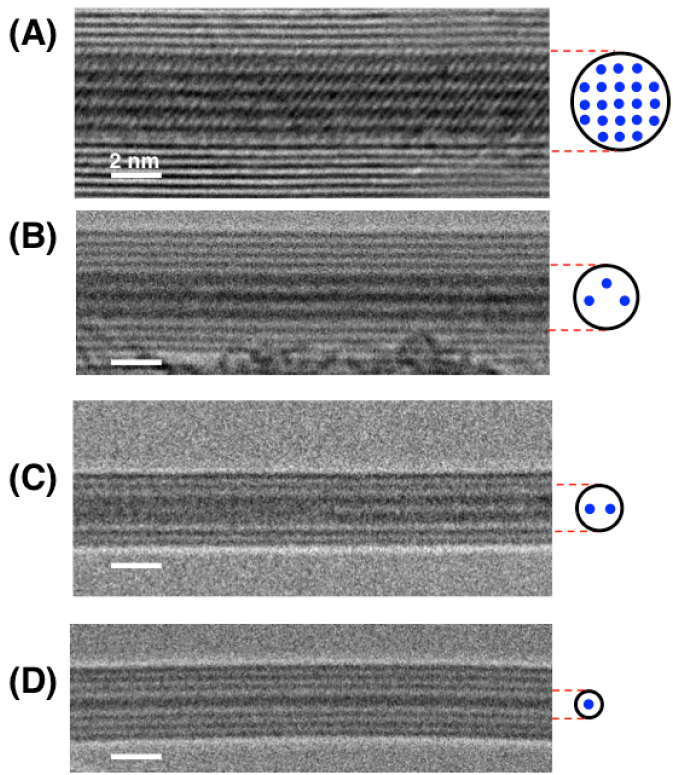

**Fig. 1.** Isolation of 1-D TMT materials down to single-chain limit. High-resolution TEM images of many (A), triple (B), double (C) and single chain (D) of prototypical TMT

NbSe$_3$ encapsulated within CNTs (A, B and C) or a BNNT (D). The simplified cross-sectional schematics show different numbers of chains encapsulated in tubes with different inner diameters. In A, atoms appear bright (over-focus); in B, C and D atoms appear dark (under-focus). The nanotubes serve as nano-reaction chambers to grow the isolated TMT chains, and simultaneously protect them from environmental degradation.

Figure 1 shows high-resolution TEM images of NbSe$_3$ chains encapsulated within nanotubes. The structure of numerous (~20) chains encapsulated by a CNT of inner diameter 3.86 nm (Fig. 1A) resembles that of the bulk crystal with signature one-dimensional (untwisted) chains oriented along the axis of the nanotube. By employing nanotubes with smaller inner diameter, fewer parallel NbSe$_3$ chains are isolated, strictly by geometrical constraint, within the cross-section of the tube. Figs. 1B, 1C and 1D show respectively triple, double, and single-chain NbSe$_3$ encapsulated within CNTs (B and C) or BNNTs (D) with successively smaller inner diameters of 2.49 nm, 1.87 nm, and 1.21 nm. This demonstrates, rather remarkably, that isolated single chains of TMTs can indeed be stabilized. We note that the unit cell of bulk NbSe$_3$ contains 6 chains (*16*), so even the 3-chain specimen is well below the single unit cell limit.

Quantitative chemical analysis for encapsulated TMT chains, employing energy dispersive spectroscopy (Fig. S1), yields 75.65 ± 7.57 at% Se and 24.35 ± 4.26 at% Nb, consistent with a stoichiometry NbSe$_3$.

Importantly, although the NbSe$_3$ chains in the few-chain limit encapsulated within nanotubes have the same stoichiometry and a similar local trigonal-prismatic atomic structure as for the bulk material, the chains do *not* precisely adopt the internal configuration found in bulk crystals. Rather, the chains are dramatically twisted, supporting a static helical torsional wave. This is true even for an isolated, single chain. For double or triple chains, the strands additionally twist around each other (forming double or triple helices), much like double-helix DNA or the strands in a multi-wire steel cable. Fig. 2A shows an aberration-corrected phase-contrast TEM (AC-TEM) image of a single NbSe$_3$ chain inside a double-walled CNT. The atomic model and the corresponding TEM simulation by the multi-slice method are also shown. The experimental image, model, and simulated image confirm the alternating orientations of the chain, i.e. the twisting of the chain. The wavelength of the associated static torsional wave, i.e. the distance for a full $2\pi$ rotation, is approximately 41nm.

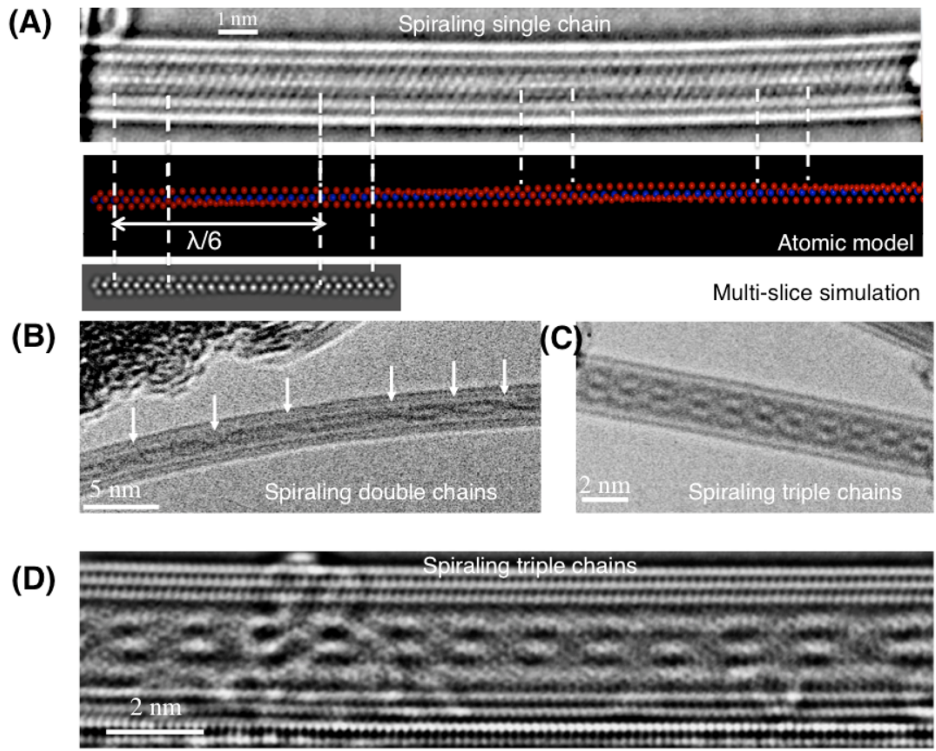

**Fig. 2.** Torsional waves of TMT chains. 2A shows an atomic-resolution phase-contrast TEM image of a spiraling single chain of NbSe$_3$ isolated inside a double-walled CNT. The structure of the single chain contains repeated patterns corresponding to different

projected orientations along its length (the dashed vertical lines delineate the repeated patterns). As illustrated in the atomic model (red: Se, blue: Nb) and multi-slice simulation, the single chain spirals with 60 degree-rotation ($2\pi/6$ or $\lambda/6$) every ~6.8 nm. The torsional waves are manifested in higher-order chains (double- and triple-chains) as displayed in Fig. 2B-D. Fig. 2B is a HR-TEM image of spiraling double-chains of $NbSe_3$ within a BNNT. The structure contains several aperiodic twisting nodes as indicated by the white arrows. Fig. 2C is a HR-TEM image, and Fig. 2D is a typical AC-TEM image, of spiraling triple-chains. Triple chains of $NbSe_3$ exhibit long range twisting with well-ordered torsional wavelengths. In A and D, atoms appear bright (over-focus), while in B and C atoms appear dark (under-focus).

Figs. 2B and 2C show additional TEM images of the spiraling behavior of double- and triple-chain specimens, respectively. For the double chain (here encapsulated within a BNNT), the additional twisting is not strictly periodic; there are significant regions where the two strands run parallel without spiraling about each other. On the other hand, for triple chains, we invariably find that the three chains are consistently tightly twisted around each other in a triple-helix fashion. Fig. 2D shows an AC-TEM image of such a triple chain configuration within a CNT. For triple chains, we find a spiraling node-node distance ranging from 1.45 to 1.85 nm within CNTs and 1.90 to 2.30 nm within BNNTs (Fig. S2B) (in the simplest interpretation the full wavelength of the torsional wave is here 6 times the node-node distance, notwithstanding additional complexities of on-chain twisting).

We comment briefly on observed dynamics of the torsional waves. Stimulation from the TEM imaging electron beam often causes the chains to bodily transport axially along the core of the tube. In addition, the wave itself can propagate along the TMT. As we show below, charging of the chains is key to the torsional wave instability. For CNTs, charging of the chains comes primarily from electron transfer from the CNT to the chain, while for BNNTs, the insulating nature of the host tube amplifies charging effects from the TEM beam (either from the beam current directly or indirectly from radiolysis processes of the BN shell or hydrocarbon contaminants nearby) (*17*, *18*) and leads to dramatic *in-situ* twisting/untwisting of the chains (as witnessed in Fig. 2B).

To explore the underlying physics of the above systems, we perform first-principles calculations based on pseudopotential density functional theory.(*19*) We first investigate the atomic and electronic structure of single-chain NbSe$_i$ isolated in vacuum. We construct three initial candidate structures for the chain using the atomic positions of the three different types of chains comprising the bulk solid (*20*) (see Fig. S5). The atomic positions for the candidate structures are fully relaxed by minimizing the total energy. All three candidates relax the same atomic structure (Fig. 3A), whose corresponding band structure is shown in Fig. 3B with two bands ($\Psi_i$ and $\Psi_2$) crossing the Fermi energy.

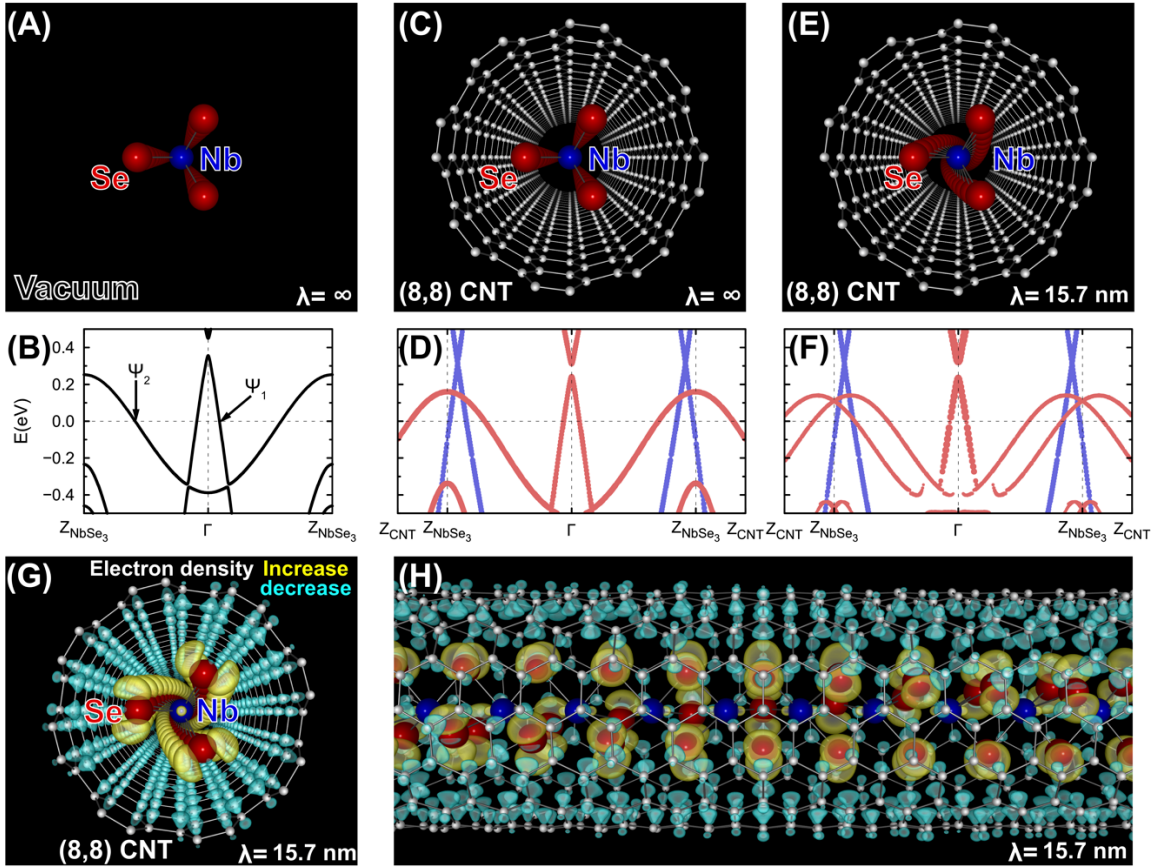

**Fig. 3.** Calculated atomic and electronic structures of single-chain NbSe$_3$. A and B show respectively the atomic and electronic band structures of untwisted single-chain NbSe$_3$ isolated in vacuum; C and D are for untwisted single-chain NbSe$_3$ encapsulated inside an (8,8) CNT, and E and F are for twisted single-chain NbSe$_3$ with λ=15.7 nm encapsulated inside the CNT. In the axial views of the atomic structures blue, red and white spheres represent Nb, Se and C atoms, respectively. In the band structures, the Fermi energy is set to zero and marked with a horizontal dashed line. In (D, F) the band structures represented by red and blue dots are projected onto the chain and CNT, respectively, and unfolded with respect to the first Brillouin zone of the unit cells of the untwisted chain and CNT, respectively, where the zone boundaries for the chain and CNT are denoted as Z$_{NbSe3}$ and Z$_{CNT}$, respectively. Figs. 3G-H display electron density transferred from the CNT to the chain with λ=15.7 nm [(G) axial and (H) lateral views]. Iso-surfaces for increased and decreased values are shaded in yellow and cyan, respectively

We next investigate the atomic and electronic structures for the untwisted single-chain $NbSe_3$ encapsulated inside an (8,8) CNT (indices chosen for convenience), Fig. 3C. The separately relaxed atomic positions of single-chain $NbSe_3$ isolated in vacuum, and those of the empty CNT, are used. Further relaxation is not performed. We calculate the binding energy $E_b$ of a single-chain $NbSe_3$ (Fig. S6), which is defined as $E_b = E_{NbSe3}+E_{CNT}-E_{NbSe3/CNT}$, where $E_{NbSe3}$, $E_{CNT}$, and $E_{NbSe3/CNT}$ are the total energies of separated single-chain $NbSe_3$ and CNT isolated in vacuum, and the joint system of single-chain $NbSe_3$ encapsulated inside the CNT, respectively. The calculated binding energy of the chain is 1.36 eV / $NbSe_3$ formula unit (f. u.). This large binding energy accounts for the stability of single-chain $NbSe_3$ encapsulated inside CNTs. Fig. 3D shows the electronic band structure of the chain inside the CNT. Confinement does not alter the states near the Fermi energy significantly, except for the charge transfer. Charge (0.23 e / f. u., i.e. 0.08 e / Se atom) is transferred from the CNT to $NbSe_3$ chain(Fig. S7), driven by the work function difference. We find no significant amount of covalent bonding between the chain and CNT (Figs. 3G-H, S7E and S8), which explains the high mobility of the chain inside the CNT.

Motivated by the experimentally observed torsional wave in single-chain $NbSe_3$, we investigate the atomic and electronic structures for the twisted single-chain encapsulated inside a CNT with a variable torsional wavelength λ. Figs. 3E-F show the atomic and electronic structures of the twisted single-chain with λ=15.7 nm. The torsional wave shifts $\Psi_2$ by $\pm$ 6π/λ, while $\Psi_1$ is not affected significantly. The torsional wave does not change the binding energy and the charge transfer significantly, and the calculated binding energies are 1.35-1.36 eV / f. u. with an electron transfer of 0.23 e/f. u. for all the calculated λs as for the untwisted chain.

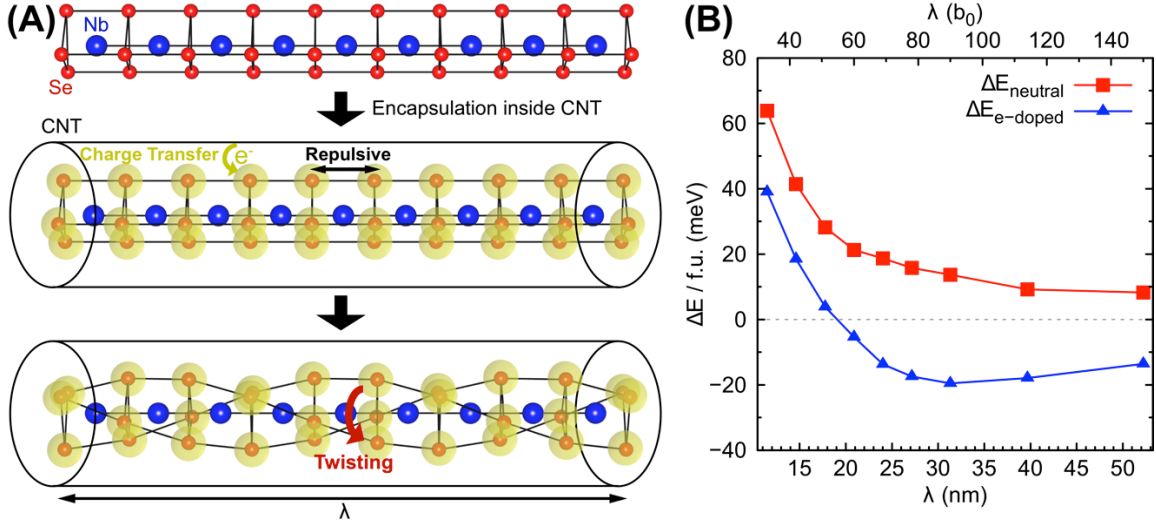

**Fig. 4.** Charge-induced Torsional Waves (CTW) in single-chain NbSe₃. Fig. 4A schematically shows the mechanism of CTW formation in single-chain NbSe₃ inside a CNT. In Fig. 4B total energies for neutral and electron-doped single-chain NbSe₃ isolated in vacuum are plotted as functions of λ, where E(∞) is set to zero and marked by a horizontal dashed line and $b_0 = 3.4805$ Å is the distance between the adjacent Nb atoms. Negative ΔE means a CTW is favored.

What drives the torsional wave? The energetics have two main contributions. The torsional wave increases the elastic energy by twisting the orbital configuration of the Nb atoms, but this is offset by a reduction in Coulomb energy between negatively charged Se atoms. To quantify these effects, we calculate the total energies of a twisted single-chain NbSe₃ isolated in vacuum with and without additional electron doping. To obtain the energy increase by twisting, we calculate the total energy difference $\Delta E_{neutral}$ of a twisted single-chain isolated in vacuum as a function of λ, defined as $\Delta E_{neutral}(\lambda) = E_{NbSe3}(\lambda) -$ $E_{NbSe3}(\infty)$, where $E_{NbSe3}(\infty)$ is the total energy of the untwisted single-chain isolated in vacuum. As shown in Fig. 4B, $\Delta E_{neutral}$ increases as $\lambda$ decreases. We also obtain the energy difference $\Delta E_{e\text{-doped}}$ for an electron-doped single-chain isolated in vacuum as a function of $\lambda$ by performing the same calculation with additional electron doping, where 0.23 e/f. u. is added to match the encapsulated situation. For $\lambda > 20$ nm, $\Delta E_{e\text{-doped}}$ is negative, and the wave distortion is favorable.

We note that within a device configuration, it should be possible to further control the charge transfer to the TMT chain(s), allowing external control of the torsional wave and thereby its optical/electronic transport properties.

**Acknowledgement**

**Funding:**

This work was primarily funded by the U.S. Department of Energy, Office of Science, Office of Basic Energy Sciences, Materials Sciences and Engineering Division, under Contract No. DE-AC02-05-CH11231 within the sp2-Bonded Materials Program (KC2207) which provided for synthesis of the chains, TEM structural characterization, and theoretical modeling.   The elemental mapping work was funded by the U.S. Department of Energy, Office of Science, Office of Basic Energy Sciences, Materials Sciences and Engineering Division, under Contract No. DE-AC02-05-CH11231 within the van der Waals Heterostructures Program (KCWF16).   Work at the Molecular Foundry (TEAM 0.5 characterization) was supported by the Office of Science, Office of Basic Energy Sciences, of the U.S. Department of Energy under Contract No. DE-AC02-05-CH11231. Support was also provided by the National Science Foundation under Grant No. DMR-1206512 which provided for preparation of opened nanotubes.

**Authors contributions:** T.P. and A.Z. conceived the idea. T.P., P.S. and S.O. synthesized the materials. T.P. and C.K. conducted TEM studies. S.O. performed DFT calculations. A.Z. and M.L.C. supervised the project. All authors contributed to the discussion of the results and writing of the manuscript.

**Competing interest:** Authors have no competing interest.

**Data and materials availability:** All data are available in the manuscript or in the supplementary materials.

**Supplementary Materials:**

Materials and Methods

Supplementary Text

Figs. S1to S8

References *(13, 21-29)*

<div align="center">

# Supplementary Materials for

## Torsional instability in the single-chain limit of a transition metal trichalcogenide

</div>

Thang Pham, Sehoon Oh, Patrick Stetz, Seita Onishi, Christian Kisielowski, Marvin L. Cohen, and Alex Zettl*

*Corresponding author: azettl@berkeley.edu

**This PDF file includes:**

Materials and Methods

Supplementary Text

Figs. S1to S8

## 1. Materials and Methods

### 1.1. Materials

**Carbon nanotubes (CNTs) and boron nitride nanotubes (BNNTs)**

CNTs are purchased from CheapTubes (90% SW-DW CNTs), while BNNTs are synthesized in-house by the Extended Pressure Inductively Coupled (EPIC) plasma method.(*21*) The nanotubes are oxidized in air to open the end caps (510 °C in 15 minutes for CNTs, and 800 °C in 60 minutes for BNNTs).

**NbSe₃ encapsulated nanotube synthesis**: 42.0 milligrams of Nb, 130.0 milligrams of Se powders (99.999 % purity from Sigma Aldrich) and 10-20 milligrams of cap-opened

CNTs/BNNTs are mixed and put into a 1/2-inch diameter quartz ampoule 20-22 centimeters long, which is evacuated and sealed. The ampoule is kept at 690 °C in a box furnace for 5-9 days.

## 1.2. Methods

**Transmission Electron Microscopy (TEM) characterization**

The as-synthesized materials are dispersed in iso-propanol by means of bath-sonication for 10-30 minutes and then dropped cast onto TEM grids for TEM characterization. TEM investigation is performed using different electron microscopes: JEOL 2010 (80 kV) for high-resolution imaging (HR-TEM), Titan-X (80 and 120 kV) for energy dispersive spectroscopy, TEAM 0.5 (80 kV, $C_S$ of -15 μm) for low-dose atomic-resolution imaging (AC-TEM).(*22*) Exit-Wave Reconstruction (EWR) is performed using 80-100 images in a focal series. Each image is taken with approximately 150-500 e$^-$/Å$^2$.s dosage to minimize the beam effect on material structure integrity and movement. Structural models are constructed using CrystalKit, and image simulation is conducted by the MacTempas software package.(*23*)

**Pseudopotential Density Functional Theory (DFT) calculations**

We use the generalized gradient approximation (*24*), norm-conserving pseudopotentials (*25*), and localized pseudo-atomic orbitals for the wave function expansion as implemented in the SIESTA code (*26*). The spin-orbit interaction is considered using fully relativistic j-dependent pseudopotentials (*27*) in the l-dependent fully-separable nonlocal form using additional Kleinman-Bylander-type projectors.(*28*, *29*) The van der Waals interaction is evaluated using the DFT-D2 correction. Dipole corrections are included to reduce the fictitious interactions between chains generated by the periodic boundary condition in our supercell approach.

## 2. Supplementary Text

**Chemical composition analysis**:

The chemical composition of as-synthesized $NbSe_3$ chains is studied and confirmed by energy dispersive spectroscopy (EDS). Fig. S1 shows a high angle annular dark field (HAADF) scanning transmission electron microscopy (STEM) image of a representative $NbSe_3$ sample, and its corresponding C, Nb and Se elemental mappings. For visualization purpose, the HAADF STEM image is adjusted to highlight the enclosed $NbSe_3$ (encapsulated CNT walls have much lower brightness). The chemical maps of both Nb and Se show very good spatial matching with the overlay dark-field STEM image. The energy dispersive spectrum of the sample is also shown. Peaks assigned to L-edges and K-edges of Nb and Se are clearly observed. The elemental quantification is based on Nb and Se K-edges with more well-defined x-ray scattering cross-sections. The quantification yields atomic ratios 75.65 ± 7.57 at% Se and 24.35 ± 4.26 at% Nb. The measurements over 20 $NbSe_3$ samples with different sizes within either CNTs or BNNTs show statistically 75.11 ± 7.67 at% Se and 24.89 ± 4.69 at% Nb. The chemical analysis confirms that the as-synthesized structures, even down to the sub-unit cell limit, still maintains a consistent atomic stoichiometry Nb:Se = 1:3 as found in bulk crystals.

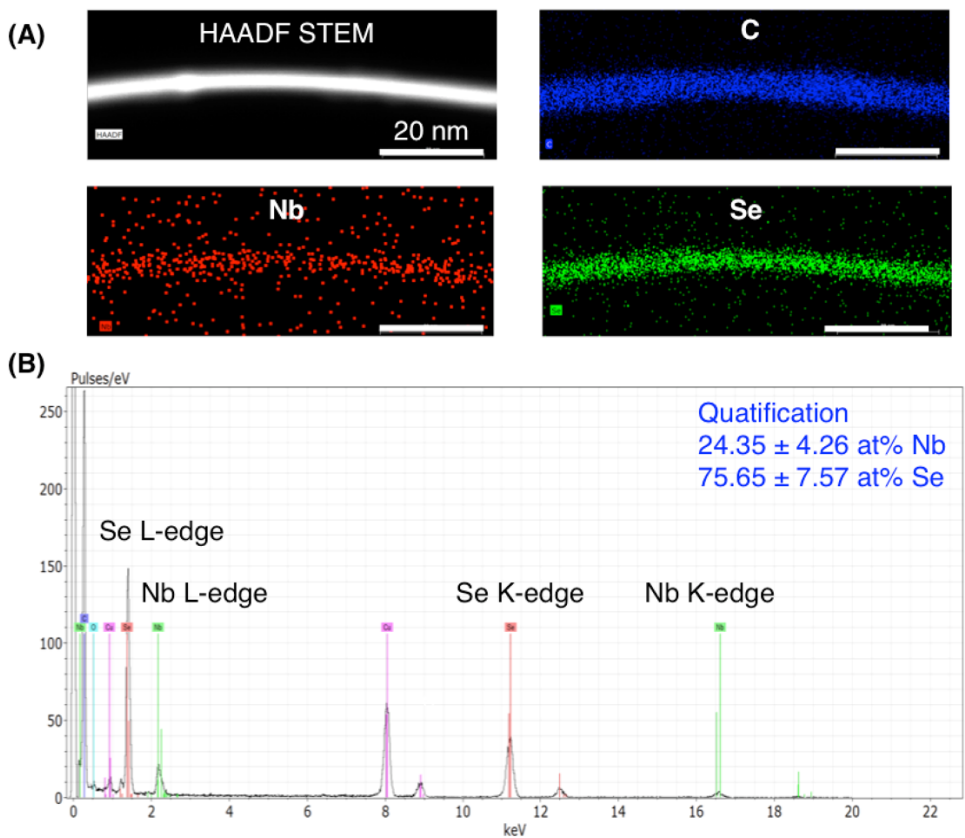

**Fig. S1**. Chemical composition analysis by energy dispersive spectroscopy (EDS). (A) High angle annular dark-field scanning TEM (HAADF STEM) of few-chains of NbSe₃ (the contrast is tuned to highlight the encapsulated NbSe₃ chains), and elemental mapping of C, Nb, and Se respectively. The chemical maps of both Nb and Se show very good spatial matching with the overlay DF-STEM image. (B) EDS spectrum showing peaks of Nb and Se L- and K-edges (Cu signals are from the TEM grid). The quantification based on the Nb and Se K-edges yields the atomic ratio of approximately 3 Se : 1 Nb.

**Nanotube inner diameter dependence**

The filling factor, which is defined as the fraction of tubes having some evidence of NbSe₃ grown inside, is more than 60 %. The length of NbSe₃ chains can be as long as several micrometers. Fig. S2a shows the distribution of inner diameters of both kinds of nanotubes (CNTs and BNNTs), which have single-, double-, and triple-chains of NbSe$_3$ formed within. For the same number of chains, the enclosed inner diameters of both CNTs and BNNTs are very similar, namely approximately 1.1 nm for single chain, 1.9 nm for double chains and 2.5 nm for triple chains. The trend shows a linear dependence between the number of NbSe$_3$ chains and the nanotube inner diameters. This demonstates that the number of NbSe$_3$ chains that form within a nanotube can be "dialed in" by simply using nanotubes with the proper inner diameter.

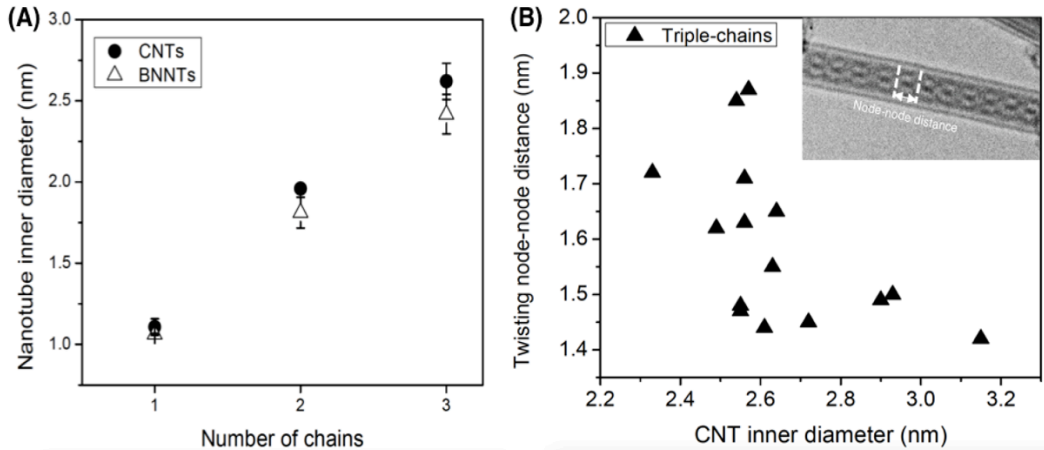

**Fig. S2.** Nanotube inner diameter dependence. (A) The relationship between the nanotube inner diameter and the number of isolated chains for single-, double- and triple-chains. (B) The dependence of CNT inner diameter and twisting node-node distance in the case of triple-chain NbSe$_3$.

Intriguingly triple chains of NbSe$_3$ almost always twist periodically along their length: of twenty triple-chain NbSe$_3$ samples surveyed, only two showed evidence for any appreciable straight length portions. Fig. S3b shows the measured twisting node-node distance in spiraling triple-chains and the corresponding CNT inner diameter over many samples. Interestingly, the twisting node-node distance in CNTs ranges from 1.45 to 1.85 nm with the center value of 1.70 nm, whereas for BNNTs (not shown here) the node – node distance ranges from 1.9 nm to 2.3 nm with the most frequent value being approximately 2.2 nm. The discrepancy in twisting periodicity in triple-chain $NbSe_3$ encapsulated in two different kinds of nanotubes may stem from two factors. The first is the difference in inner diameter of tubes filled with three-chain $NbSe_3$. As shown in Fig. S2a the inner diameters of the corresponding BNNTs, 2.6 nm on average, are slightly larger than those of CNTs, 2.4 nm on average. The second factor is the difference in charge transfer mechanism in the two host nanotubes. While in semiconducting/metallic carbon nanotubes the charge (electron) can transfer directly from the carbon sheath to the chains (as discussed in details in the main text), in insulating BNNTs the charge transfer occurs from the electron beam in the TEM, or indirectly from secondary electrons coming from radiolysis process of the BN shell, or of nearby hydrocarbon contaminants.

**Many-chain $NbSe_3$**

Figure S3 is a representative AC-TEM image of many-chain $NbSe_3$, along with corresponding selected area electron diffraction, atomic model, and TEM multi-slice simulation. In electron diffraction, the zone axis is [12-1] and the main diffractions are indexed as (101) (corresponding to 6.90 ± 0.1 Å, the inter-chain distance in this projection), (012) (corresponding to 3.2 ± 0.1 Å, the distance between the dark lobes along the chains), and (113) using the lattice parameters of bulk $NbSe_3$.(*16*)

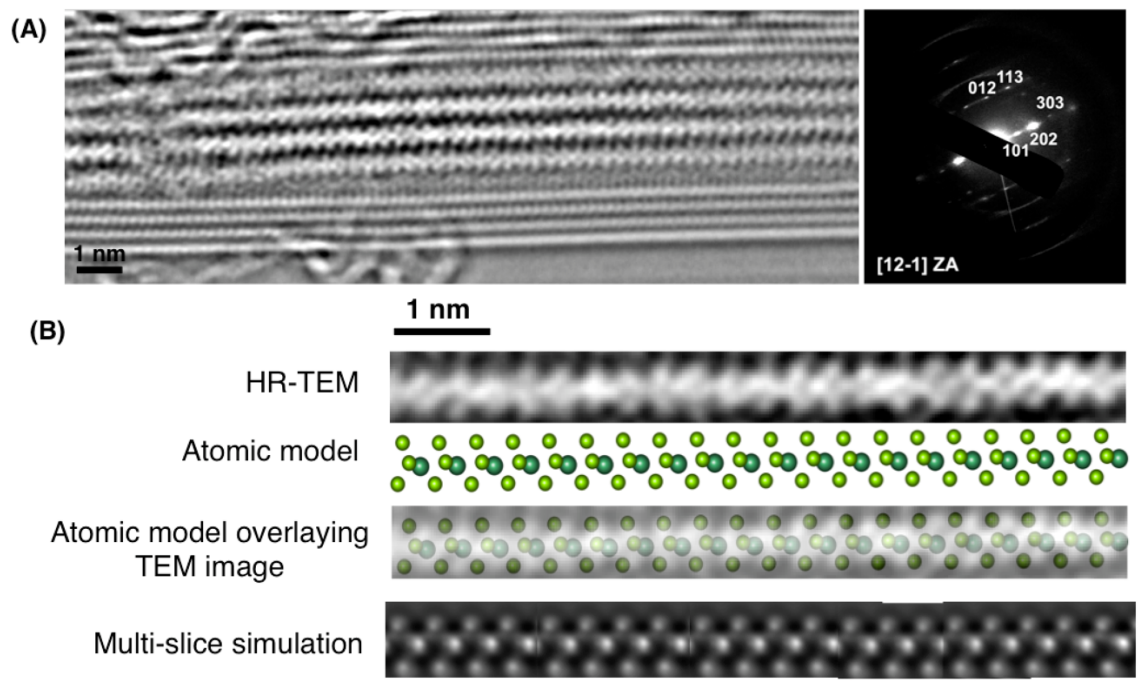

**Fig. S3.** Many chain NbSe₃. (A) An atomic-resolution phase-contrast TEM image of many (~20) parallel chains NbSe₃ and the corresponding selected area electron diffraction in [12-1] zone axis. The main indexed diffraction spots are (101), (202), (303), (102) and (113). (B) One chain is cut out with the atomic model (Nb: green, Se: light green) and the corresponding multi-slice simulation.

**Single-chain NbSe₃ isolated in vacuum**

We investigate theoretically the atomic and electronic structure of a single-chain NbSe₃ isolated in vacuum. We construct three initial candidate structures for the chain using the atomic positions of the three different types of chains composing the bulk solid and assume a vacuum region of 40 Å x 40 Å perpendicular to the chain as shown in Fig. S5. From the constructed initial candidate structures, the atomic positions are fully relaxed by minimizing the total energy. All three candidates relax to the same atomic structure, shown in Fig. S5G. As the structure relaxes, Se atoms on the same plane perpendicular to the chain tend to form equilateral triangles. Figure S5H shows the electronic band structure of the relaxed structure. There are two bands crossing the Fermi energy ($\Psi_1$ and $\Psi_2$). One band($\Psi_1$) crossing 0.08 Å$^{-1}$ away from $\Gamma$ point mainly consists of the $d_z^2$ orbitals of Nb atoms and the $p_z$ orbitals of Se atoms, and the other band($\Psi_2$) crossing 0.83 Å$^{-1}$ away from $\Gamma$ point mostly consists of the $p_\Theta$ orbital of Se atoms, where $\Theta$ represent the azimuthal direction. The magnitudes of two nesting vectors related to the two bands are 0.16 and 1.06 Å$^{-1}$, respectively.

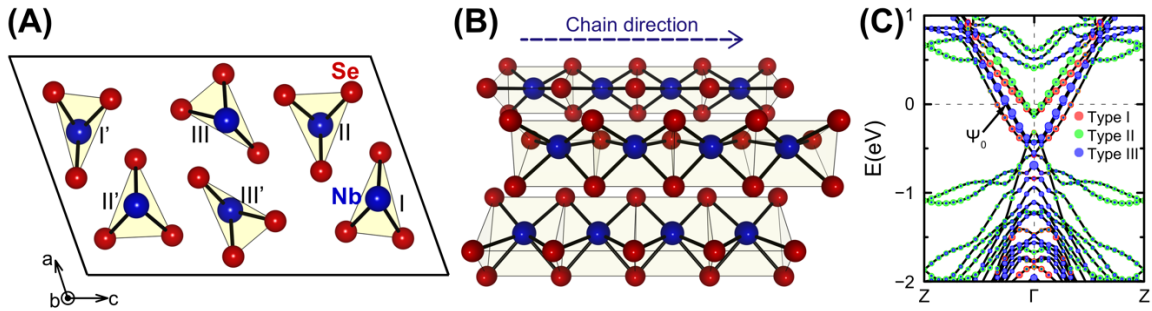

**Fig. S4.** Atomic and electronic structures of NbSe$_3$ bulk solid. A and B show the atomic structure of NbSe$_3$ bulk solid [(A) top and (B) side views], where blue and red spheres represent Nb and Se atoms, respectively. In C the calculated electronic band structure is shown along the chain direction including the $\Gamma$ point. The Fermi energy is set to zero and marked with a horizontal dashed line. The size of red, green and blue dots represents the contribution of type I, type II, and type III chains, respectively. The band crossing the Fermi energy 0.22 Å$^{-1}$ away from $\Gamma$ point is denoted as $\Psi_0$, which is associated with the higher temperature charge density wave with the nesting vector q = 0.44 Å$^{-1}$. As shown in C, $\Psi_0$ mostly comes from the type III chain.

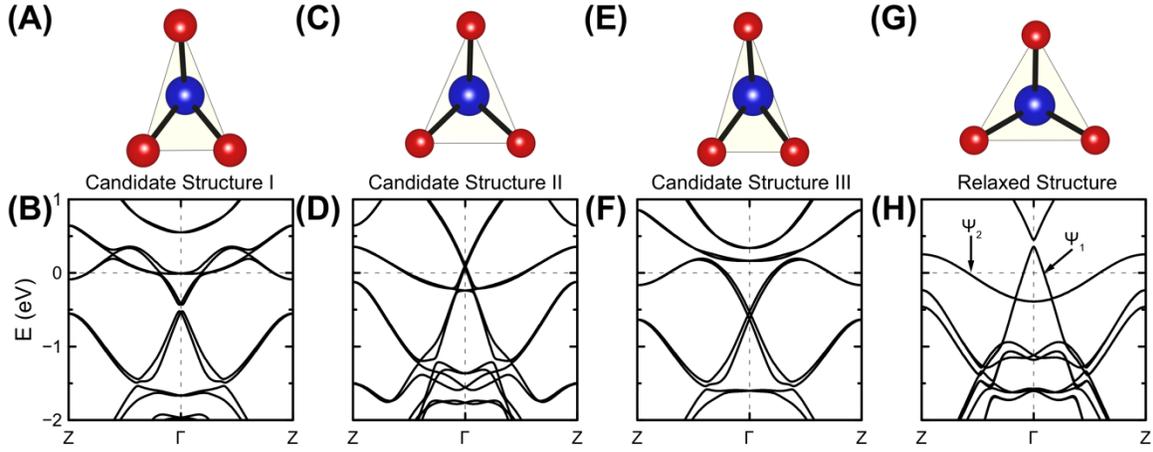

**Fig. S5.** Calculated atomic and electronic structures of single-chain NbSe$_3$ isolated in vacuum. In A-F the atomic and electronic structures of three candidate structures for single-chain NbSe$_3$ isolated in vacuum are shown. Candidate structures I, II, and III are respectively constructed using the atomic positions of type I, II, and III chains comprising the bulk solid as denoted in Fig. S4A. A and B show respectively the atomic and electronic band structures of candidate structure I; C and D are for candidate structure II, and E and F are for candidate structure III. The relaxed atomic structure and corresponding electronic band structure are shown in G and H, respectively.

**Binding energy of single-chain NbSe$_3$ encapsulated in CNT**

We obtain the binding energy $E_b$ of single-chain NbSe$_3$ encapsulated in a CNT, which is defined as $E_b = E_{NbSe3} + E_{CNT} - E_{NbSe3/CNT}$, where $E_{NbSe3}$, $E_{CNT}$, and $E_{NbSe3/CNT}$ are the total energies of separated single-chain NbSe$_3$ isolated in vacuum, the CNT, and the joint system of single-chain NbSe$_3$ encapsulated inside the CNT, respectively. Figure S6 shows the binding energy of the chain inside CNTs as a function of CNT diameter.

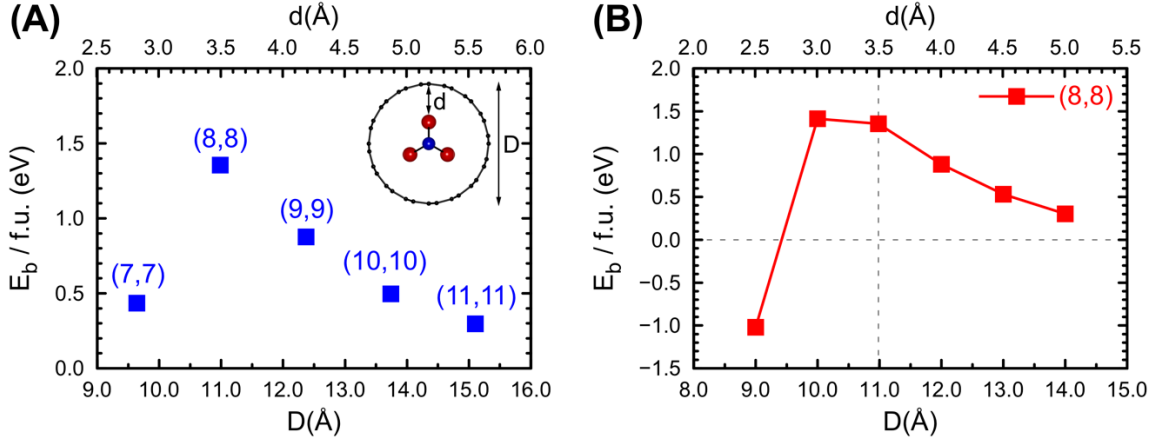

**Fig. S6.** Binding energy of single-chain $NbSe_3$ encapsulated within a CNT. In (A) binding energies $E_b$ of the chain inside various armchair CNTs are plotted with the atomic structure in the inset. In (B) $E_b$ of the chain inside an (8,8) CNT with radial compression is plotted as a function of CNT diameter D, where the diameter of the uncompressed CNT is marked with a vertical dashed line.

**Charge transfer in single-chain $NbSe_3$ encapsulated in CNT**

We obtain the number of electrons transferred from the CNT to a single-chain $NbSe_3$ from Mulliken population analysis, 0.23 e / $NbSe_3$ formula unit (f. u.), i. e. 0.08 e / Se atom. To confirm, we calculate the charge density difference and band structure of the single-chain isolated in vacuum with 0.23 e / f. u. added to the chain. The charge density difference is obtained by subtracting the electron density of the neutral chain from that of the electron-doped chain. The calculated charge difference and the band structure match well with the encapsulated situation as shown in Figs. S7C-F.

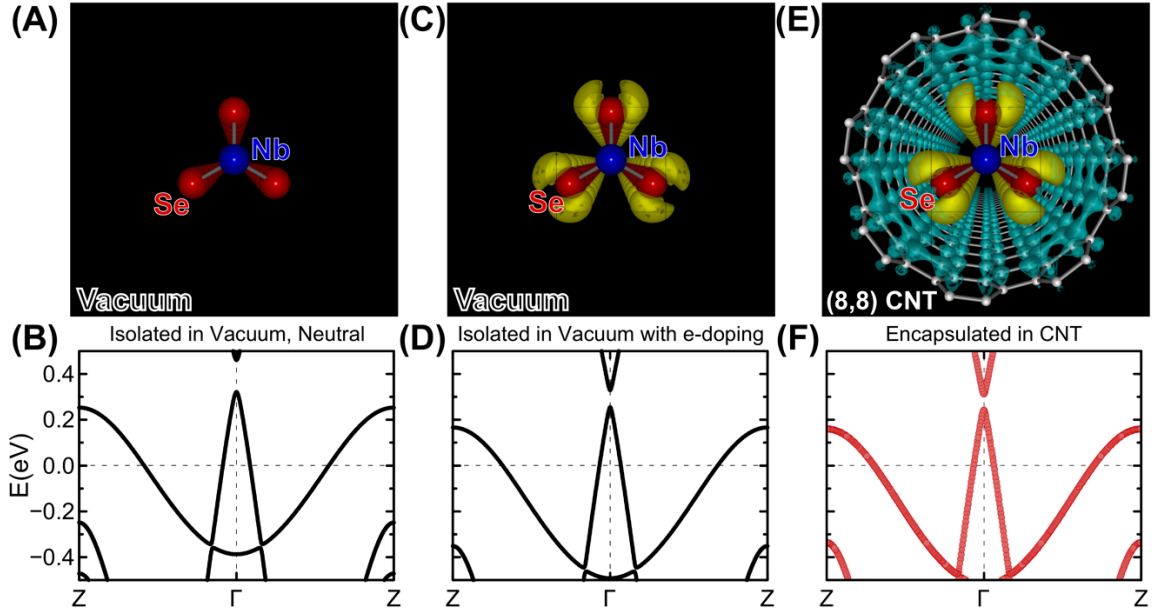

**Fig. S7.** Calculation for charge transfer in single-chain NbSe₃ encapsulated in a CNT. A and B show respectively the atomic and electronic band structures of neutral single-chain NbSe₃ isolated in vacuum; C and D are for an electron-doped single-chain isolated in vacuum, and E and F for single-chain NbSe3 encapsulated inside an (8,8) CNT. In (C,D) 0.23 e/f.u. is added to match the encapsulated situation. In (C) the iso-surface for the added electron is shaded in yellow. In (E) the electron density transferred from CNT to the chain is also shown, where iso-surfaces for increased and decreased values are shaded in yellow and cyan, respectively. In (F) the band structures are projected onto the chain and unfolded with respect to the first Brillouin zone of the unit cell of the chain.

## Covalent bonding between single-chain NbSe₃ and CNT

In order to check the existence of covalent bonding between single-chain NbSe₃ and the CNT, we carefully check the charge density of the chain inside the CNT and the charge density difference obtained by subtracting the electron densities of the chain isolated in vacuum, and an empty CNT, from that of the joint system. Very little charge is found in the regions between the Se atoms and the CNT as shown in Figs. S7E and S8. As a check, we analyze all of the occupied states below the Fermi energy. No significant amount of covalent bonding between single-chain NbSe$_3$ and CNT is found in all the investigated states.

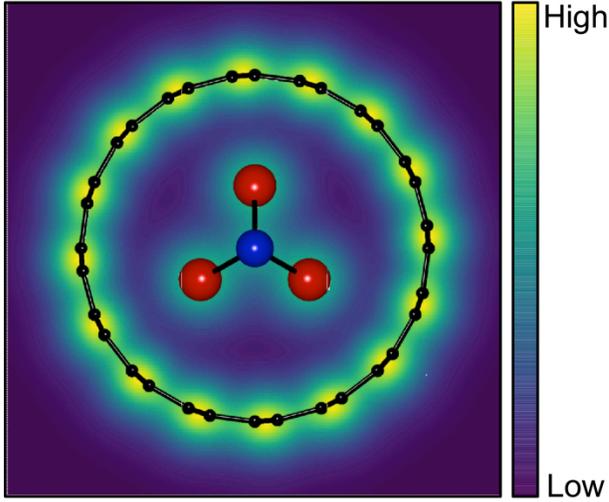

**Fig. S8.** Charge density of single-chain NbSe$_3$ encapsulated in CNT. Averaged electron density for a single-chain NbSe$_3$ encapsulated in a (8,8) CNT is plotted, superposed with the atomic structure. The averaged electron density n(r,Θ) is defined as $n(r,\theta) = \frac{1}{L}\int_0^L dz\, n(r,\theta,z)$, where r, Θ, and z are the radial, azimuthal, and axial coordinates, respectively, n(r,Θ,z) is the electron density at (r,Θ,z) and L is the length of unitcell vector in the axial direction.

**Twisted single-chain encapsulated inside CNT**

We investigate the atomic and electronic structures for the twisted single-chain encapsulated inside a CNT with a variable torsional wavelength λ. Because the electronic structure calculation for encapsulated chain with long wavelength λ is very intensive, we investigate the electronic structures with $\lambda = 15.7$ nm as shown in Figs. 3F-H of the main paper, which shows the effects of the CTW on the electronic structure.